\documentclass[manuscript]{aastex}
\usepackage{emulateapj5}
\newcommand{\co}{\rm CO}
\newcommand{\hh}{{\rm H}_2}

\newcommand{\kps}{\,\textstyle\rm{km~s}^{-1}}

\newcommand{\cobm}{{\rm CO(2-1)}}
\newcommand{\coam}{{\rm CO(1-0)}}
\newcommand{\tto}{{\rm 2-1}}
\newcommand{\otz}{{\rm 1-0}}

\newcommand{\cob}{CO(2$\rightarrow$1)~}

\newcommand{\yr}{\,\textstyle\rm{yr}}

\newcommand{\msun}{\,M_{\odot}}
\newcommand{\lsun}{\,L_{\odot}}

\newcommand{\jy}{\,\textstyle\rm{Jy}}

\newcommand{\Kkpspc}{\,\rm{K}\,\rm{km~s}^{-1}\,{\rm pc}^{2}}

\slugcomment{Accepted ApJL 02 May 2008}

\shorttitle{Molecular Gas in GOODS\,J123634.53+621241.3}
\shortauthors{Frayer et al.}

\begin{document}

\title{Molecular Gas in the {\scriptsize z}=1.2 Ultraluminous
Merger GOODS\,J123634.53+621241.3}

\author{David T. Frayer\altaffilmark{1}, Jin Koda\altaffilmark{2},
Alexandra Pope\altaffilmark{3,4}, Minh T. Huynh\altaffilmark{1},
Ranga-Ram Chary\altaffilmark{5}, Douglas Scott\altaffilmark{6}, Mark
Dickinson\altaffilmark{4}, Douglas C.-J. Bock\altaffilmark{7}, John
M. Carpenter\altaffilmark{2}, David Hawkins\altaffilmark{8}, Mark
Hodges\altaffilmark{8}, James W. Lamb\altaffilmark{8}, Richard
L. Plambeck\altaffilmark{9}, Marc W. Pound\altaffilmark{10}, Stephen
L. Scott\altaffilmark{8}, Nicholas Z. Scoville\altaffilmark{2}, and
David P. Woody\altaffilmark{8}}

\altaffiltext{1}{Infrared Processing and Analysis Center, California
Institute of Technology 100-22, Pasadena, CA 91125, USA;
frayer@ipac.caltech.edu}

%\altaffiltext{2}{NASA Herschel Science Center, California
%Institute of Technology 100-22, Pasadena, CA 91125, USA}

\altaffiltext{2}{Astronomy Department, California Institute of Technology
105--24, Pasadena, CA  91125, USA} 

\altaffiltext{3}{{\em Spitzer} Fellow}

\altaffiltext{4}{National Optical Astronomy Observatory, 950 N. Cherry
Ave., Tucson, AZ, 85719, USA}

\altaffiltext{5}{{\em Spitzer} Science Center, California Institute of
Technology 220-6, Pasadena, CA 91125, USA}

\altaffiltext{6}{Department of Physics \& Astronomy, University of
British Columbia, Vancouver, BC, V6T 1Z1, Canada}

\altaffiltext{7}{Combined Array for Research in Millimeter-wave
Astronomy, P.O Box 968, Big Pine, CA 93513, USA}

\altaffiltext{8}{Owens Valley Radio Observatory, California Institute
of Technology, Big Pine, CA 93513, USA}

\altaffiltext{9}{Department of Astronomy and Radio Astronomy
Laboratory, University of California, Berkeley, CA 94720, USA}

\altaffiltext{10}{Department of Astronomy, University of Maryland,
College Park, MD 20742, USA}

\begin{abstract}

We report the detection of \cob emission from the $z=1.2$
ultraluminous infrared galaxy (ULIRG) GOODS\,J123634.53+621241.3 (also
known as the sub-millimeter galaxy GN26).  These observations
represent the first discovery of high-redshift CO emission using the
new Combined Array for Research in Millimeter-Wave Astronomy (CARMA).
Of all high-redshift ($z>1$) galaxies within the GOODS-North field,
this source has the largest far-infrared (FIR) flux observed in the
{\em Spitzer} 70\,$\mu$m and 160\,$\mu$m bands.  The CO redshift
confirms the optical identification of the source, and the bright \cob
line suggests the presence of a large molecular gas reservoir of about
$7\times 10^{10} M_{\sun}$.  The infrared-to-CO luminosity ratio of
$L({\rm IR})/L^{\prime}(\co) = 80\pm 30 \lsun (\Kkpspc)^{-1}$ is
slightly smaller than the average ratio found in local ULIRGs and
high-redshift sub-millimeter galaxies.  The short star-formation time
scale of about 70\,Myr is consistent with a starburst associated with
the merger event and is much shorter than the time scales for spiral
galaxies and estimates made for high-redshift galaxies selected on the
basis of their $B-z$ and $z-K$ colors.

\end{abstract}


\section{Introduction}

Observations of molecular gas are fundamental to our understanding of
galaxy evolution by providing measurements of the material from which
stars form.  The discovery of sub-millimeter galaxies
\citep[SMGs,][]{sma97, hug98, bar98} enabled the ability to measure
the molecular CO properties of the most luminous infrared sources at
high-redshift \citep{fra98, fra99, ner03, gre05, tac06, tac08}.  These
observations have shown that the CO properties of the SMGs are similar
to local ultraluminous infrared galaxies (ULIRGs, $L_{\rm IR} >
10^{12} \lsun$), and their number densities indicate that ULIRGs are
1000 times more common at $z \sim 2$ than locally \citep{cha05}.  The
importance of ULIRGs at high-redshift has been reinforced by deep
{\em Spitzer} mid-infrared (MIR) surveys at 24\,$\mu$m
\citep[e.g.,][]{cha04, lag04}.  At $z\sim 1$--2, the majority of
star-formation and AGN activity occurs within dust-rich luminous
infrared galaxies \citep{lef05, pap06, cap07}.

Although the bulk of the infrared luminosity of galaxies arises in the
rest-frame far-infrared band (FIR, 40--120\,$\mu$m), this band has
been mostly unexplored at high-redshift.  The majority of
high-redshift ULIRGs have been identified by 24\,$\mu$m and sub-mm/mm
surveys.  These surveys yield highly uncertain bolometric corrections
for the infrared luminosity.  Deep surveys at 70\,$\mu$m and
160\,$\mu$m with {\em Spitzer} enable the direct identification of
luminous sources within the FIR band.  In this Letter we report on CO
observations of the brightest FIR source at $z>1$ in GOODS-North.  The
source GOODS\,J123634.53+621241.3 (also known as SMG GN26) was
identified as being unusually bright at 70\,\&\,160\,$\mu$m
\citep{fra06,huy07} and for having strong emission from polycyclic
aromatic hydrocarbon (PAH) molecules \citep{pop08}.  At optical
wavelengths, GN26 has a merger-like morphology with a disturbed core
and evidence of tidal tails on larger scales.  Although it has a large
FIR flux, it is among the faintest 850~$\mu$m sources detected to date
\citep{bor03, pop06}.  A cosmology of $h_{70}\equiv {\rm H}_0
(70\kps\,{\rm Mpc}^{-1})^{-1} = 1$, $\Omega_{\rm M}=0.3$, and
$\Omega_{\Lambda}=0.7$ is assumed throughout this paper.

\section{Observations}

The \cob observations of GN26 were taken between 2007 April 14 -- June
28 with the Combined Array for Research in Millimeter-Wave Astronomy
(CARMA) in the low-resolution D-configuration, resulting in a total of
26.4 hours of on-source data.  The adopted phase center was the IRAC
position of $\alpha$(J2000)=$12^{\rm h} 36^{\rm m}34\fs5$,
$\delta$(J2000)=$+62\arcdeg 12\arcmin 41\arcsec$ \citep{pop06}.  The
CO line was observed using a digital correlator configured with two
adjacent $13\times31.25$\,MHz bands centered on 103.89274 GHz in the
upper side-band, corresponding to \cob emission at the reported
optical redshift of $z=1.219$ \citep{coh96}.  The nearby quasar
J1153+495 ($14\arcdeg$ from GN26) was observed every 15 minutes for
amplitude and phase calibration, and the bright quasars 3C\,273,
3C\,345, and J0927+390 were used for passband calibration and
pointing.  We estimate an uncertainty of 14\% for the absolute
calibration scale based on the observations of 3C\,273 and 3C\,345.

The data were reduced and calibrated using the Multichannel Image
Reconstruction, Image Analysis, and Display (MIRIAD) software package
\citep{sau95}.  An interactive UV-plotting tool was used to visualize
and flag spurious visibilities per baseline (105 baselines in total).
Figure~1 shows the ``dirty'' (no cleaning) natural-weighted integrated
\cob map made by averaging over the 10 channels showing CO emission at
the phase center.  No improvement in image quality was made using
clean algorithms since the side-lobes are weaker than the noise due to
the large number of baselines.  The data are consistent with an
unresolved point source.  The weak tail extending eastward is not
currently significant (Fig.~1).  Higher resolution data with higher
signal-to-noise would be needed to measure the CO morphology of the
source.
 
\section{Results}

The total \cob line flux of $3.45 \pm 0.46\jy\kps$ was derived from
the integrated CO map ($7.5\sigma$, Fig.~1).  The CO position and the
limit to the CO source size of $<4\arcsec$ (33\,kpc) were derived from
a ``robust''-weighted image (ROBUST$=0$).  The derived position is
consistent with previous optical and radio positions.  The CO emission
peaks at the central location of the merger event.  With the current
sensitivity and resolution, there is no evidence for CO emission from
the nearby companion galaxy GOODS J123634.69+621243.7 at $z=1.225$
\citep{cow04}, which is about $3\arcsec$ north-east of GN26.

Figure~2 shows the \cob spectrum at the peak of the CO image.  The
spectrum was Hanning-smoothed for display purposes, but a Gaussian fit
of the un-smoothed data was used to derive the CO redshift and line
width (Table~1).  The CO redshift of $z(\co)=1.2234\pm0.0007$ is
slightly red-ward ($600\kps$) of the reported optical redshift of
$z=1.219$ \citep{coh96,cow04}, which is not unusual for dust-obscured
ultra-luminous systems \citep{fra99, gre05}.  \cite{wir04} observed
the galaxy with the Keck DEIMOS spectrograph but did not report a
redshift.  However, visual inspection of the spectrum using the Team
Keck Treasury Redshift Survey on-line database shows an emission line
consistent with [OII] $\lambda3727$\AA\ at a redshift of
$z=1.224\pm0.001$, which agrees well with the measured CO redshift.
In addition, the CO redshift is consistent with the redshift based on
the PAH lines ($z=1.23\pm0.01$) observed with {\em Spitzer} IRS
\citep{pop08}.  Based on its sub-mm and radio flux densities
\citep[S850\,$= 2.2$\,mJy; S1.4\,GHz\,$=0.19$\,mJy,][]{pop06}, the
continuum level at 3\,mm is expected to be less than about 0.05\,mJy.
As expected, no continuum was detected from the co-addition of the
lower and upper side-band line-free channels; S(3mm)$ < 0.66$\,mJy
($3\sigma$).  Since the 3~mm continuum level is negligible in
comparison to the strong CO line, no continuum was subtracted from the
data.

The observed \cob line flux implies an intrinsic CO line luminosity of
$L^{\prime}[\cobm] = (6.8\pm1.8) \times 10^{10} \Kkpspc$ \citep[see
formulae in][]{sv05}, which is a factor of 150--200 times larger than
that for our Galaxy.  The CO luminosity is related to the mass of
molecular gas (including He) by $M(\hh)/L^{\prime}(\co) = \alpha$.
Given that \cite{tac08} have found that the conversion factor $\alpha$
for the SMGs is similar to that derived for local ULIRGs, we have
adopted the average ULIRG value for GN26.  Assuming $\alpha(\otz) =
0.8 \msun(\Kkpspc)^{-1}$ and
$L^{\prime}[\cobm]/L^{\prime}[\coam]\simeq 0.8$ as found for local
ULIRGs \citep{ds98}, we adopt $\alpha(\tto) = 1 \msun(\Kkpspc)^{-1}$;
i.e., $\alpha(\tto) = \alpha(\otz)
(L^{\prime}[\cobm]/L^{\prime}[\coam])^{-1}$.  For $\alpha(\tto) = 1$,
the molecular gas mass for GN26 is $M(\hh) \simeq 7\times 10^{10}
M_{\sun}$.

The baryonic gas fraction provides an estimate of the evolutionary
state of the system [$\mu \equiv M_{\rm gas}/(M_{\rm gas}+M_{\rm
stars})]$ \citep[e.g.,][]{fb97}.  Massive local spirals have gas
fractions of $\mu \simeq 0.05$--0.1 \citep{ys91}, while younger
gas-rich systems have higher gas fractions and more evolved systems
have lower gas fractions.  To derive the stellar mass and the gas
fraction for GN26, we used the \cite{bc03} population synthesis models
to fit the multi-wavelength (U through z-band and {\em Spitzer} IRAC
3-6$\mu$m) photometry.  We assumed solar metallicities, a Salpeter
IMF, and star-formation histories ranging from a single instantaneous
burst to various $e$-folding timescales, including constant
star-formation.  We estimate an age of the stellar population of
1.3~Gyr and a total stellar mass of
(2.1$\pm$0.6)$\times$10$^{11}$\,M$_{\sun}$ for an e-folding timescale
of 1.0~Gyr.  Neglecting H{\sc I} ($M_{gas} \simeq M(\hh)$), the
implied gas fraction is $\mu = 0.25\pm0.10$.

The dynamical mass is highly uncertain given the lack of resolution of
the CO data and the uncertain kinematics associated with the merger
event.  However, the CO size of $\theta({\rm FWHM})<4\arcsec$
(diameter $< 33$\,kpc) can provide a crude constraint on the dynamical
mass.  For a wide range of mass distributions, the dynamical mass can
be approximated by $M_{\rm dyn}\approx R(\Delta V/[2\sin(i)])^2/G$,
where $\Delta V$ is the observed FWHM line width, $G$ is the
gravitational constant, and $i$ is the inclination.  If the intrinsic
axial ratio is about 1, the observed optical morphology implies an
inclination of $i\simeq 40\arcdeg$.  For a radius of $R<16.5$\,kpc, a
CO line width of $560\kps$, and $i\simeq 40\arcdeg$, the dynamical
mass is $M_{\rm dyn} < 7\times 10^{11} \msun$.  This result implies a
gas-fraction of larger than 10\% which is consistent with the estimate
based on the stellar mass.

We derive a FIR (42.5--122.5\,$\mu$m) luminosity of $L({\rm FIR}) =
(4.0\pm1.2) \times 10^{12} L_{\odot}$ and a total infrared luminosity
(8--1000\,$\mu$m)\footnote{We adopt an IR definition of
  8--1000\,$\mu$m and the {\em IRAS} FIR definition of
  42.5--122.5\,$\mu$m throughout this paper.} of $L({\rm IR}) =
(5.6\pm1.7) \times 10^{12} L_{\odot}$, using the 70, 160, and
850\,$\mu$m flux densities \citep{pop06, huy07} and the IRS spectrum
\citep{pop08}.  A single-temperature greybody with $T_{d}= 41\pm3$\,K
and a dust-emissivity index of $\beta \simeq 2$ fit the data.  Similar
luminosities have been previously derived based on local spectral
energy distribution (SED) templates \citep{huy07,pop08}, but a simple
greybody provides a better fit to the FIR data for this system.

\section{Discussion}

GN26 has a disturbed, merger-like morphology (Fig. 1) similar to local
ULIRGs and many high-redshift ULIRGs/SMGs.  The {\em Spitzer} mid-infrared
IRS spectrum \citep{pop08}, the infrared-to-radio flux density ratio
\citep{huy07}, and the X-ray emission \citep{ale05} are all consistent
with GN26 being predominantly powered by star formation.  The large
\cob flux is consistent with this view.

We derive a total star-formation rate of SFR\,$\simeq 950 \msun
\yr^{-1}$, using SFR($\msun \yr^{-1}) = 1.7\times 10^{-10} L({\rm
IR})/\lsun$ which assumes a Salpeter IMF \citep{ken98}.  The $L({\rm
IR})/L^{\prime}(\co)$ ratio provides an indication of the intensity of
star-formation and star-formation efficiency.  Strong starbursts and
ULIRGs tend to show high IR-to-CO luminosity ratios of $\ga 100$,
while local spiral galaxies have lower values of about 10--50
\citep[e.g.,][]{sv05}.  For GN26, $L({\rm IR})/L^{\prime}(\co) = 80\pm
30\,\lsun (\Kkpspc)^{-1}$.  An average ratio of 350 has been reported
for high-redshift CO sources \citep{sv05}.  However, previous $L({\rm
IR})$ estimates for the SMGs assume a temperature of 40\,K
\citep{gre05}, while multi-wavelength data suggest a lower average
dust temperature of about 35\,K for the SMG population \citep{cha05,
kov06, pop06, pop08, huy07}, implying a factor of two decrease in
their luminosities.  If adopting an average temperature of 35\,K for
the SMGs and placing the data on the same infrared luminosity scale,
we find that both the high-redshift SMGs \citep{gre05} and local
ULIRGs \citep{sol97} have a similar average value of $L({\rm
IR})/L^{\prime}(\co) \simeq 200\pm 100 \lsun (\Kkpspc)^{-1}$.  The
value for GN26 is on the low end of the range of values found for the
more distant SMGs and local ULIRGs.

In comparison, the recent CO observations of the massive $z\sim 2$ BzK
galaxies \citep[galaxies selected on the basis of their $B-z$ and
$z-K$ colors,][]{dad07} suggest a similar ratio of $L({\rm
IR})/L^{\prime}(\co) = 60\pm30\lsun (\Kkpspc)^{-1}$ \citep{dad08}.
Although the IR-to-CO luminosity ratios for GN26 and the BzKs are
similar, the BzKs are suspected to have much longer star-formation
time scales.  For the adopted CO-to-H$_2$ conversion factor
($\alpha[\tto] \simeq 1)$, the estimated star-formation (gas
consumption) time scale of $M(\hh)/{\rm SFR}= 70$\,Myr for GN26 is
consistent with a merger-driven starburst and the short time scales
estimated previously for starbursts in local ULIRGs and distant SMGs
\citep{sv05, tac08}. In contrast, the BzK are thought to have much
longer star-formation time scales of order $\ga 400$\,Myr based on
their estimated duty cycle \citep{dad07}.  The gas consumption time
scales for the BzKs based on the CO data would be consistent with
these results if the Galactic CO-to-H$_2$ conversion factor is adopted
for the BzKs \citep[e.g., $\alpha(\otz) = 4.8
\msun(\Kkpspc)^{-1}$,][]{san91}.  Within this paradigm, BzK galaxies
would be analogous to local spirals with enhanced SFRs and long
star-formation time scales, while the SMGs would be analogous to local
ULIRGs with short time scales.  However, such generalizations may be
an over simplification.  High-redshift galaxies such as GN26 and the
BzKs may have a wide range of star-formation time scales between the
intense short-lived starbursts found in local ULIRGs of order 10\,Myr
and the long time scales of order 1\,Gyr found for local spirals.
Future high-resolution CO observations are needed to constrain the
CO-to-H$_2$ conversion factors and the associated gas consumption time
scales before definitive conclusions may be drawn.

GN26 is not representative of most SMGs studied to date.  It has a
relatively low redshift, is faint at 850\,$\mu$m, and is warmer than
other SMGs detected at $z\sim 1$ \citep{cha05}.  It also has the
largest observed S(CO)/S850 ratio of any SMG to date by a factor of 5
(in part due to the K-correction of this ratio).  Previous
high-redshift CO surveys have concentrated on the brightest
850\,$\mu$m sources which are typically at $z>2$.  For sources at
$z\sim1$, the {\em Spitzer} 70 and 160\,$\mu$m surveys and future
Herschel surveys which probe the peak of the FIR SED may be more
effective at identifying the brightest CO sources than the current
850\,$\mu$m surveys.  The relative importance of low-redshift SMGs
\citep[$z\sim1$,][]{wan06,wan08}, ``typical'' SMGs \citep[$2\la z
\la3$,][]{bla02, cha05}, and the most distant SMGs
\cite[$z>3$,][]{dun04, wan07, you07, you08, dan08} is an active topic
of discussion.  \cite{wal08} have recently argued for two distinct
populations of SMGs: ULIRGs at $z\sim1$ and more luminous ULIRGs at
$z\sim$2--3.  Comparing the CO properties of SMGs as a function of
redshift and luminosity may help to test the hypothesis of multiple
SMG populations.

\section{Concluding Remarks}

GN26 is a merger-driven starburst with a molecular gas mass of about
$7\times 10^{10} M_{\sun}$.  Although it is the brightest FIR
(70\,$\mu$m and 160\,$\mu$m) high-redshift ($z>1$) source in
GOODS-North, it has not been previously observed in CO given that it
is among the faintest 850\,$\mu$m sources.  The CO properties of GN26
are consistent with local ULIRGs and the high-redshift SMGs. The
derived gas fraction of $0.25\pm0.10$ based on the stellar and
molecular gas mass estimates suggests that about 75\% of the available
baryonic mass has already been converted into stars.  Furthermore, at
the current star-formation rate, the molecular gas will be depleted
within $\la 100$~Myr.  The short gas consumption time scale for GN26
is in contrast to long time scales found for local spiral galaxies and
estimates made for the high-redshift BzKs.

These observations represent the first discovery of high-redshift CO
using CARMA.  Planned improvements of the CARMA receivers will allow
for CO observations of significant samples of distant ULIRGs
discovered by the ongoing {\em Spitzer} programs and future Herschel
surveys.

\acknowledgments

Support for CARMA construction was derived from the Gordon and Betty
Moore Foundation, the Kenneth T. and Eileen L. Norris Foundation, the
Associates of the California Institute of Technology, the states of
California, Illinois, and Maryland, and the National Science
Foundation.  Ongoing CARMA development and operations are supported by
the National Science Foundation under a cooperative agreement, and by
the CARMA partner universities.  This work is based in part on
observations made with the {\em Spitzer} Space Telescope and has made use of
the NASA/IPAC Extragalactic Database (NED), both of which are operated
by the Jet Propulsion Laboratory, California Institute of Technology
under a contract with NASA.  The optical image was obtained from the
Multi-mission Archive at the Space Telescope Science Institute (MAST).

%\clearpage

\begin{table}
\begin{center}
\caption{\cob Observational Results}
\tablewidth{300pt}
\begin{tabular}{lr}
\tableline\tableline
Parameter & Value \\
\tableline
$\alpha$(J2000)  & $12^{\rm h} 36^{\rm m}34\fs51\pm0\fs08$ \\
$\delta$(J2000) &$+62\arcdeg 12\arcmin 41\farcs2\pm0\farcs5$ \\
$z(\co)$  &$1.2234\pm0.0007$ \\
Linewidth (FWHM)& $560\pm90\kps$\\
$S(\co)$\tablenotemark{a}& $3.45\pm0.93\jy\kps$ \\
$L(\co)$\tablenotemark&  $(2.7\pm0.7)\times 10^{7} h_{70}^{-2}
L_{\sun}$ \\
$L^{\prime}(\co)$\tablenotemark& $(6.8\pm1.8)\times 10^{10}
h_{70}^{-2}\Kkpspc$\\
$M(\hh)$\tablenotemark{b} & $7\times10^{10} h_{70}^{-2} M_{\sun}$\\
\tableline
\end{tabular}
\tablenotetext{a}{Observed total \cob line flux.  Uncertainty
includes the 14\% systematic calibration uncertainty and the
13\% ($7.5\sigma$ detection) random noise.}
\tablenotetext{b}{Estimated assuming $\alpha(\tto) = 1 \msun
(\Kkpspc)^{-1}$.}
\end{center}
\end{table}

\clearpage

\begin{figure}
\epsscale{.60}
\plotone{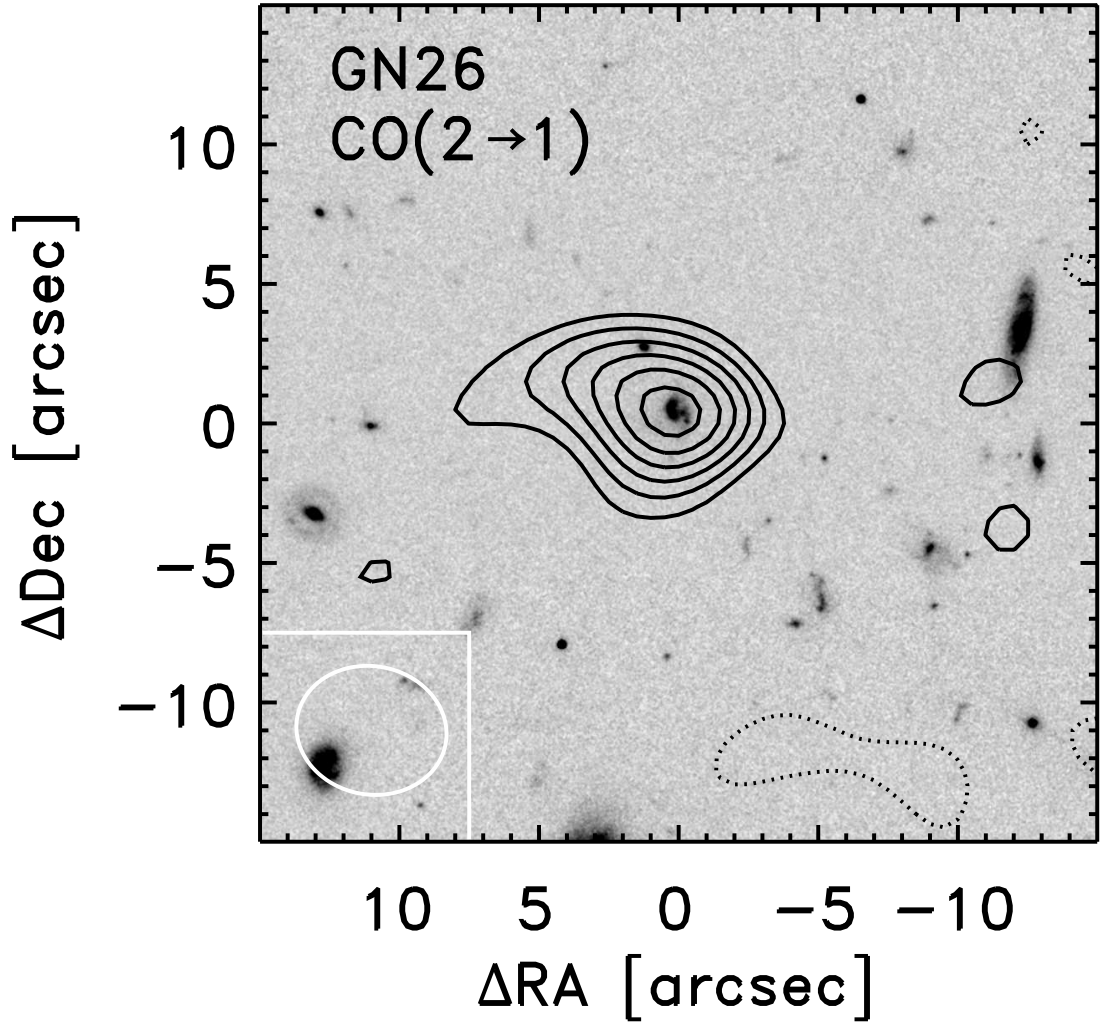}
\vspace*{3cm}
\caption{The integrated \cob map of GN26 averaged over 312.5\,MHz
  ($904\kps$) overlaid on an optical {\em HST}-ACS $z$-band image
  (grey-scale).  The $1\sigma$ error is $0.46\jy\kps$/beam, and the
  contour levels are $1\sigma\times$($-$2,2,3,4,5,6,7); positive
  contours are solid lines and negative contours are dotted lines.
  The synthesized beam size (white ellipse) for the observations is
  shown in the lower left ($5\farcs4\times4\farcs6, {\rm
  PA}=80\arcdeg$).}

\end{figure}

\clearpage

\begin{figure}
\epsscale{0.9}
\plotone{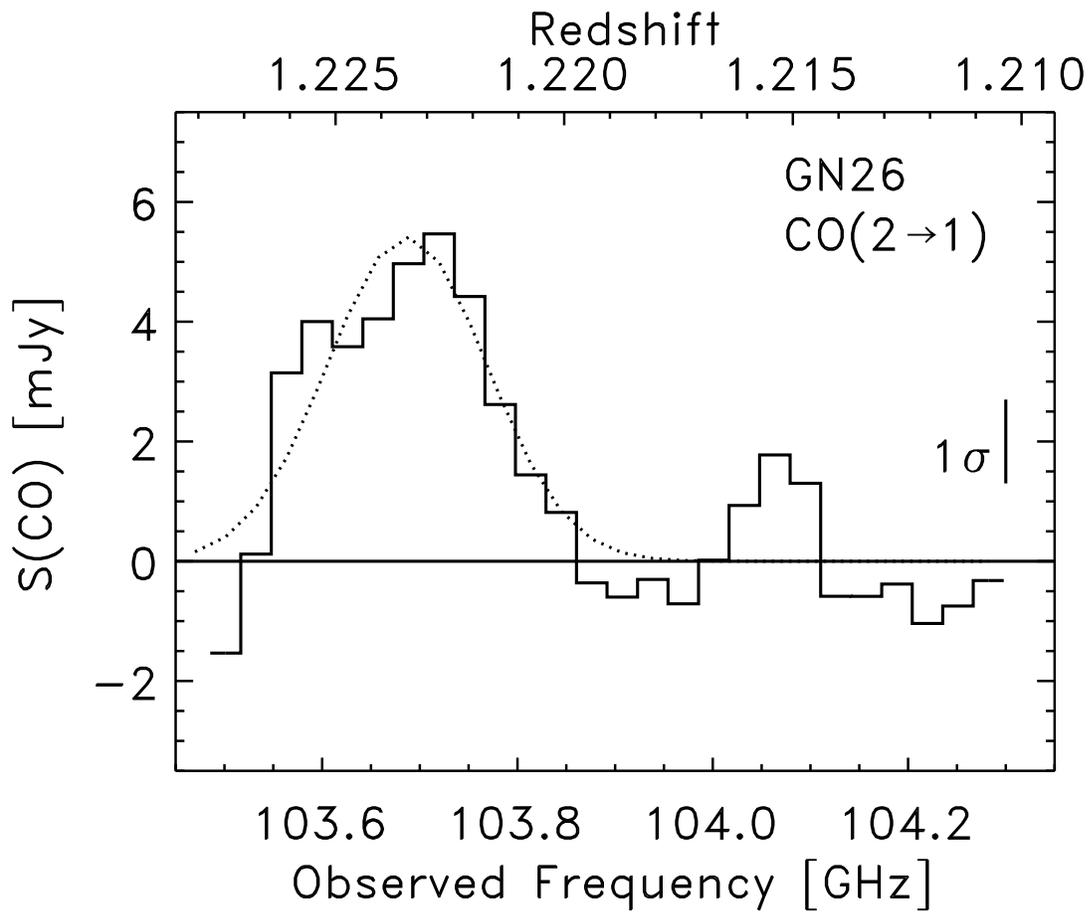}
\caption{ The \cob spectrum for GN26.  The spectrum has been
Hanning-smoothed.  The channel increment is 31.25\,MHz ($90.4\kps$),
and the $1\sigma$ error bar per smoothed channel is shown to the
right.  The dotted-line shows a Gaussian fit of the line profile for
the unsmoothed data.}

\end{figure}

\end{document}